\documentclass[%
 aps, prl, reprint,
 superscriptaddress,
 longbibliography, nopreprintnumbers, noeprint,
 amsmath, amssymb, floatfix
]{revtex4-2}

\usepackage{multirow}
\usepackage{graphicx}
\usepackage{dcolumn}
\usepackage{bm}
\usepackage{dsfont}
\usepackage{hyperref}
\usepackage{physics}
\usepackage[usenames,dvipsnames]{xcolor}
\usepackage{bbold}
\usepackage[normalem]{ulem} 
\usepackage[inline]{enumitem}
\usepackage{qcircuit} 
\usepackage{tikz}
\usepackage{amssymb} 
\usepackage[caption=false]{subfig}
\usepackage{algorithm} 
\usepackage{algpseudocode} 
\usepackage{mathrsfs}
\usepackage{circuitikz} 
\DeclareMathAlphabet{\mathpzc}{OT1}{pzc}{m}{it}
\usepackage{booktabs} 

\usepackage{verbatim}

\newcommand{\figref}[1]{Fig.~\ref{#1}}

\renewcommand{\eqref}[1]{Eq.~\ref{#1}}

\renewcommand{\section}[1]{\textit{#1.} --- }

\begin{document}

\title{Long-lived memory effects in the defect bath of superconducting qubits}

\author{Abhishek Agarwal}
\email{abhishek.agarwal@npl.co.uk}
\author{Masum Uddin}
\author{Shroya Vaidya}
\author{Lachlan P. Lindoy}
\author{Ehsaneh Daghigh Ahmadi}
\author{Tobias Lindstr\"om}
\author{Sebastian~E.~de~Graaf}
\affiliation{National  Physical  Laboratory,  Teddington,  TW11  0LW,  United  Kingdom}
\author{Ivan Rungger}
\email{ivan.rungger@npl.co.uk}
\affiliation{National  Physical  Laboratory,  Teddington,  TW11  0LW,  United  Kingdom}
\affiliation{Department of Computer Science, Royal Holloway, University of London, Egham, TW20 0EX, United Kingdom}

\begin{abstract}
We reveal long-lived memory effects in the defect bath of a superconducting transmon qubit through electric-field  tuning of two-level system (TLS) defects coupled to the qubit. Using a fast TLS mapping method we observe several hysteretic effects in the TLS environment with memory timescales of the order of seconds, far beyond the lifetimes of individual TLS defects. 
The observations can be explained by TLS coupling to electric field-polarised charge fluctuators in the defect bath. Our method enables detailed mapping of the dynamics of the bath's coupled microscopic degrees of freedom and the associated memory effects which can introduce temporally correlated noise. This information may be used to improve qubit-stabilisation and quantum error correction protocols.
\end{abstract}    

\maketitle

\section{Introduction}
Microscopic two-level-system (TLS) defects are a dominant source of dielectric loss and temporal instability in superconducting qubits~\cite{muller2015interacting,martinis2005decoherence,muller2019towards,de_graaf_two-level_2020,Klimov_2018,schlor2019correlating}. Their coupling to the qubit causes significant qubit relaxation~\cite{Klimov_2018,lisenfeld_electric_2019,zanuz_mitigating_2024}, while slow spectral diffusion leads to device-to-device and time-dependent variations in coherence~\cite{carroll2022dynamics,weeden_statistics_2025, zanuz_mitigating_2024, berritta_real-time_2025, berritta2026adaptivespectroscopyfasttwolevelsystem}. This instability is particularly problematic for quantum error correction (QEC), where qubit performance must remain predictable between calibrations and across repeated error-correction cycles~\cite{acharya2024quantum,f_kam_detrimental_2025}.
The temporal correlations associated with these fluctuations depend on how long the defect environment retains memory of its previous state. Recent observations of hyperpolarisation and non-Markovian relaxation and dephasing indicate long-lived internal degrees of freedom in the TLS environment~\cite{spiecker_two-level_2023,gosling_probing_2026,zhuang_non-markovian_2026, gao_non-local_2026}. However, microscopic mechanisms and timescales are not well understood.

Electric-field tuning~\cite{lisenfeld_electric_2019,lisenfeld_mapping_2026,chen2025scalablesitespecificfrequencytuning,bilmes_resolving_2020} provides a way to probe such memory effects in the TLS defect environment. 
Here we show that the defect environment can retain a memory of its electric-field history over long timescales. We perform fast, repeated maps of the qubit relaxation probability while a gate electrode in the transmon enclosure sweeps the applied electric field.
Strongly coupled TLS defects reproducibly exhibit history-dependent resonance voltages, including cases where the same TLS crosses the qubit resonance once in one sweep direction and twice in the other due to a TLS frequency jump induced by the changing field. We show that these behaviours are consistent with field-polarised charge fluctuators (CFs) that shift the local asymmetry energy of the resonant TLS to which they are coupled~\cite{meisner_probing_2018,muller2015interacting,de2021quantifying,faoro2015interacting}. Unlike two-level-fluctuators (TLF), which generally refer to thermally active low-energy defects~\cite{faoro2015interacting,de2021quantifying}, these CFs can have large tunnelling barriers~\cite{pourkabirian_nonequilibrium_2014} which reduce thermal fluctuations and give rise to the hysteretic behaviour upon electric field sweeping.

In our experiments, sweep-rate-dependent measurements reveal fluctuators with equilibration times of the order of seconds. The timescales associated with these fluctuators are remarkably long compared to typical TLS baths, where coherence times are typically measured in microseconds, with only a few recent works demonstrating relaxation times in the 10-100 ms range \cite{chen2024phononengineering, gosling_probing_2026}, and with the TLS-phonon relaxation rates typically  seen in amorphous glasses in the kHz range \cite{faoro2015interacting}. These memory effects in the defect bath can introduce long-lived temporal correlations in the qubit environment, giving rise to detrimental non-Markovian noise~\cite{agarwal_fast-tracking_2025,agarwal2023modelling,burkard_non-markovian_2009,zhuang_non-markovian_2026} and placing practical constraints on electric-field-based qubit stabilization and QEC protocols~\cite{chen2025scalablesitespecificfrequencytuning, lisenfeld_enhancing_2023, dane2025performancestabilizationhighcoherencesuperconducting,kim_error_2025}.

\section{Fast defect environment mapping}
An applied electric field  $\mathbf{E}$ shifts the transition frequency of a TLS with electric dipole moment $\mathbf{p}$ through the projection $\mathbf{p}\cdot\mathbf{E}$~\cite{lisenfeld_electric_2019}. In our experiments the applied field comes from a gate electrode which is placed approximately $1\,\text{mm}$ above the transmon qubit chip. We use a standard aluminium-on-silicon transmon that has a ground-to-excited state transition frequency $\sim 4.4\,\text{GHz}$. The qubit transition frequency is not affected significantly by the applied electric field since transmons are designed to be insensitive to offset charge~\cite{koch_charge-insensitive_2007}. However, the qubit relaxation rate, $\Gamma_1$, increases when a coupled TLS defect comes into resonance~\cite{barends_coherent_2013}. Measuring $\Gamma_1$ as a function of applied field therefore maps the electric defect environment of the qubit~\cite{barends_coherent_2013,lisenfeld_electric_2019}.
 Standard relaxation rate measurements can be slow~\cite{krantz_quantum_2019}; thus, we use a proxy metric provided by measuring the probability of the qubit remaining in its excited state, $P_e$, after a delay of time $\tau_\text{delay}$~\cite{carroll2022dynamics}. Assuming ideal preparation and measurement at zero temperature, $P_e(\tau_\text{delay}) = e^{-\Gamma_1 \tau_\text{delay}}$. To balance the trade-off between time resolution and resolvability of small changes in $\Gamma_1$, we use a delay time of $\tau_\text{delay} =10\,\mu \text{s}$ based on the typical $T_1$ time of $\sim 30\,\mu\text{s}$. Repeated measurements of $P_e$ are performed in parallel with a time-varying applied electric field, which enables measuring $P_e$ as a function of the applied field.

\begin{figure}[tb]
    \centering
    \includegraphics[width=\columnwidth]{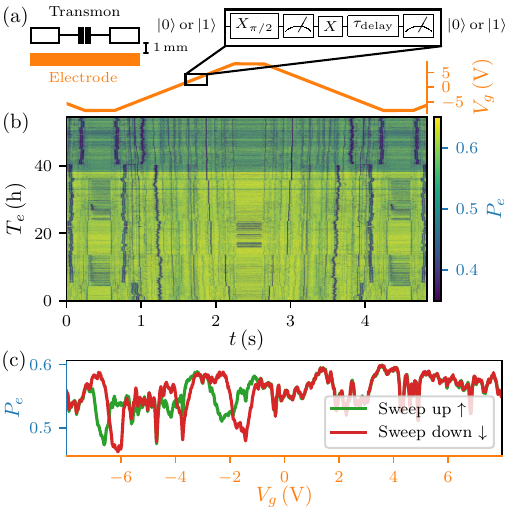}
    \caption{(a) The experimental protocol showing the circuit repeatedly applied on the qubit (black) and the waveform of the applied voltage $V_g$ via the gate electrode (orange). In the circuit diagram, the $X_{\pi/2}$ pulse followed by readout randomly projects the qubit to the $\ket{0}$ or $\ket{1}$ state. Measurements where the qubit was projected to $\ket{1}$ are discarded in post-processing~\cite{werninghaus_high-speed_2021}. The remaining circuit is a single-point $T_1$ measurement. (b) Probability of the qubit remaining excited, $P_e$, as a function of sweep time $t$ within each voltage sweep, plotted over the laboratory time $T_e$ at which each experiment was performed. The dark vertical trajectories correspond to individual TLS defects that come into resonance with the qubit. (c) The average $P_e$ over the $\sim 60\text{h}$ of the experiment plotted as a function of voltage $V_g$ for the increasing and decreasing sweep directions.}
    \label{fig:fig1}
\end{figure}

We implement the pulse sequence shown in~\figref{fig:fig1}(a), which can be run without real-time-control based active state preparation~\cite{riste2012feedbackcontrol} while also avoiding the long delays of passive resets. To evaluate $P_e (t)$ in minimal time, where $t$ is the time since the start of the individual experiment, we average over $64$ repetitions, which yields sufficient single-shot statistics to resolve the changes in $P_e$ produced by strongly coupled TLS defects. This results in millisecond-scale time resolution, comparable with some of the fastest relaxation rate measurements~\cite{berritta_real-time_2025}, albeit with larger uncertainty.
In a single voltage sweep indicated in~\figref{fig:fig1}(a), we perform $64\times4096=262144$ repetitions of the pulse sequence while the gate voltage $V_g(t)$ is ramped. This experiment is then repeated over $\sim60\,\text{h}$. 
A gate voltage is assigned to each measurement by determining the global offset of the $V_g$ waveform using the observed periodicity in $P_e$.

An example of the obtained results for $P_e$ is shown in ~\figref{fig:fig1}(b), where dips in $P_e$ corresponding to enhanced qubit relaxation near TLS resonances are seen. These persist over the elapsed laboratory time, $T_e$, while exhibiting gradual drifts, small diffusive fluctuations, and occasional large discrete jumps, in agreement with what is commonly observed for TLS fluctuations~\cite{Klimov_2018,weeden_statistics_2025,bejanin_interacting_2021}. Note that a change in the background $P_e$ occurs near $\sim38\text{h}$ which could be caused by an increase in qubit relaxation due to a source which is not tuned by the applied electric field, or changes in qubit or readout properties; it does not affect the sweep-direction comparison below.
Assuming that the voltage is approximately constant within the averaging window, the extracted $P_e(t)$ is correlated with the applied voltage $V_g(t)$ to evaluate $P_e(V_g)$.

Measurements acquired during increasing and decreasing voltage ramps are denoted by $P_{e,\uparrow}(V_g)$ and $P_{e,\downarrow}(V_g)$, respectively. The average values of these over the $\sim60\,\text{h}$ of repeated experiments are plotted in \figref{fig:fig1}(c). In the absence of memory effects in the defect environment, $P_{e,\uparrow}(V_g)$ and $P_{e,\downarrow}(V_g)$ would be identical. However, we observe significant differences between  $P_{e,\uparrow} $ and $ P_{e,\downarrow}$ (clearly visible for $V_g\lesssim0 \text{V}$ in ~\figref{fig:fig1} (c)), indicating memory effects in the defect environment~\cite{meisner_probing_2018}.

\section{Individual TLS tracking}
\begin{figure*}[htb]
    \centering
    \includegraphics[width=1\textwidth]{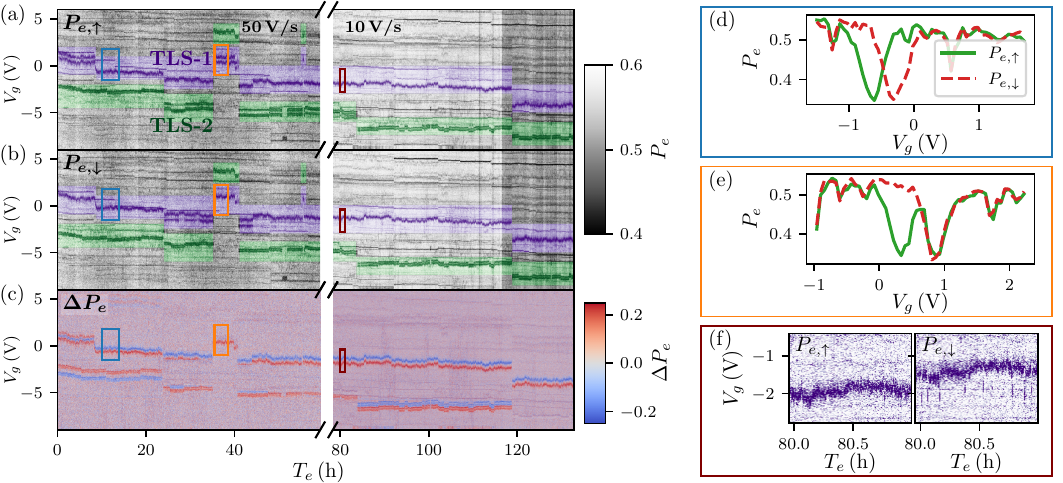}
    \caption{Two sets of repeated experiments are performed with sweep rates $50\,\text{V}\text{s}^{-1}$ ($T_e = 0-60\,\text{h}$) and $10\,\text{V}\text{s}^{-1}$ ($T_e = 80-130\,\text{h}$) and the quantities (a) $P_{e,\uparrow}$, (b) $P_{e,\downarrow}$, and (c) $\Delta P_{e} = P_{e,\uparrow}-P_{e,\downarrow}$ are plotted as a function of $V_g$ over time. The areas surrounding the two dominant strongly coupled TLS, labelled TLS-1 and TLS-2, are highlighted in purple and green, respectively, for better visibility. (d-f) Representative examples of different hysteresis effects, with the colour of the box corresponding to the regions marked in (a-c) with the same colours. Panels (d,e) show the time-averaged result during the boxed region.
    (f) Zoomed-in view of the region highlighted in the maroon boxes showing hysteresis similar to (d) but with the addition of probabilistic jumps when the voltage is decreasing (right subplot), an effect that would not be clearly visible in a histogram.} 
    \label{fig:fig2}
\end{figure*}
As a first step towards understanding the origins of the hysteretic behaviour, we analyse individual TLS defects and their evolution over time in \figref{fig:fig2}. The results are obtained from two sets of repeated experiments with different voltage ramp rates; the pulse sequence, voltage sweep range, and total sweep period are kept fixed by adjusting the hold duration between ramps. \figref{fig:fig2}(a,b) show the results for the two sweep directions, while \figref{fig:fig2}(c) reveals hysteretic effects of two dominant defects. Note that hysteresis effects are common and also visible for other defects; however, due to weaker TLS coupling strengths the corresponding features are harder to resolve and require scans with finer voltage resolution and averaging for reliable identification and analysis.

The boxes in \figref{fig:fig2}(d-f) highlight three representative forms of hysteresis that recur at different times and affect both TLS-1 and TLS-2. We now describe each of these in turn. 

\textbf{Conventional hysteresis:} In \figref{fig:fig2}(d), the TLS resonance voltage depends on the sweep direction. Similar behaviour has previously been attributed to coupling between a TLS and a slow field-polarised charge fluctuator, which causes $P_e(t)$ to depend on both the instantaneous voltage $V_g(t)$ and its prior evolution~\cite{meisner_probing_2018,lisenfeld_enhancing_2023}. Our repeated mapping shows that this hysteresis persists over many hours and is therefore not a transient artefact.

\textbf{Reappearance hysteresis:} In \figref{fig:fig2}(e), the same TLS crosses the qubit resonance twice in one sweep direction but only once in the opposite direction, producing three dips in $P_e$. Their similar widths, magnitudes, and time-dependence indicate that they originate from the same TLS. Mechanisms which might generate such behaviour are discussed in the following sections.

\textbf{Probabilistic hysteresis:} In \figref{fig:fig2}(f), the TLS generally appears at different resonance voltages for the two sweep directions but occasionally switches to the same position. These events are absent from the faster-sweep data at $T_e<60\,\text{h}$ and appear in the slower-sweep data at $T_e>80\,\text{h}$. This indicates sweep-rate-dependent switching, with the slower sweep approaching the characteristic timescale of the underlying memory process.

\section{Phenomenological mechanism for the observed hysteresis}
We now introduce a phenomenological charge-fluctuator model that accounts for these behaviours, and then use interleaved sweep-rate-dependent measurements to estimate the memory timescale.
A field-polarised charge fluctuator (CF) provides a minimal mechanism for conventional hysteresis~\cite{meisner_probing_2018, pourkabirian_nonequilibrium_2014}. Assuming a large tunnelling barrier, the two localised configurations of the CF are approximately its energy eigenstates and its asymmetry energy can be assumed to vary linearly with gate voltage, $\epsilon_\text{CF}(V_g)=\epsilon_\text{CF}(0)+\nu_\text{CF}V_g$. The large tunnelling barrier suppresses thermal and tunnelling-induced transitions between the two configurations, making the configurations metastable. Suppose that far from the symmetry point $\epsilon_\text{CF}=0$, the CF is initially in its ground state configuration.
As $V_g$ is varied and the symmetry point is crossed, the configuration corresponding to the ground state of the CF changes. However, the large tunnelling barrier prevents the CF from equilibrating to its new ground state immediately. The CF remains in its initially occupied, metastable configuration until it equilibrates to its ground state later in the sweep.
The state of the CF near the symmetry point then depends on the sweep direction, since that determines which state the CF was initially polarised to. A coupled TLS is shifted by different amounts for the two CF states, producing the conventional hysteresis in which the TLS resonance position depends on sweep direction. 
When the time spent near the symmetry point is comparable to the CF equilibration time, the same mechanism gives probabilistic hysteresis as the switching may occur between the two TLS resonance positions during some sweeps but not others.

As $V_g$ is swept beyond the symmetry point of the CF, the effective barrier preventing equilibration reduces. At a specific voltage $V_\text{th}$ the effective barrier is small enough that the CF can switch to its ground state on a fast timescale relative to the time resolution of $\sim 1\,\text{ms}$. For the two CF states, if the resonance positions of the TLS lie on opposite sides of the CF switching point, the same TLS crosses the qubit frequency twice during one sweep, producing the double resonance feature in~\figref{fig:fig2}(e). Note that this switching process can also give rise to the TLS not coming into resonance at all and going undetected if the switch moves the TLS resonance location to a voltage that has already been swept.

\begin{figure}[tb]
    \centering
    \includegraphics[width=1\columnwidth]{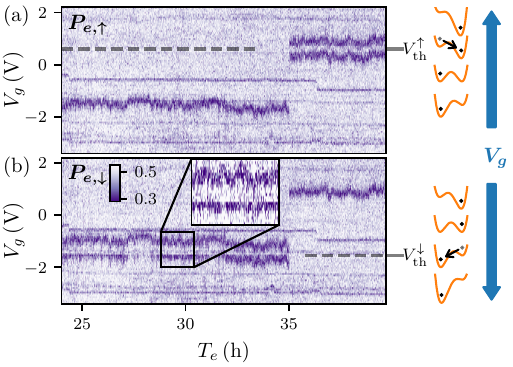}
    \caption{Zoomed-in view of the results in~\figref{fig:fig2}(a,b) showing (a) $P_{e,\uparrow}$ and (b) $P_{e,\downarrow}$. In (b), for times $<35\,\text{h}$,  TLS-1 reappears abruptly at a well-defined threshold voltage, consistent with switching of the CF. This abrupt reappearance is highlighted in the high-contrast zoomed-in inset in (b). The inferred CF thresholds $V_\text{th}^\uparrow$ and $V_\text{th}^\downarrow$ are marked on the right, together with a schematic representation of how the CF potential varies with the applied $V_g$ and causes the reappearance hysteresis.}
    \label{fig:fig3}
\end{figure}

A zoomed-in view of the reappearance hysteresis behaviour is shown in~\figref{fig:fig3}, which shows the CF switching near its two switching locations $V_\text{th}^\uparrow$ and $V_\text{th}^\downarrow$ for the two sweeping directions. A schematic for how the shape of the CF potential well changes with the field is also shown, indicating these switching locations.
This model predicts a sharp onset or termination of the TLS induced dip in $P_\mathrm{e}$ when the CF switching threshold overlaps with the TLS resonance. Such a threshold is observed in~\figref{fig:fig3}(b) between $24\,\text{h}$ and $31\,\text{h}$, providing further evidence that the reappearance is associated with a charge fluctuator switching at a well-defined electric field.
This model, therefore, can explain the various observed hysteresis effects seen on TLS-1. 
The analysis above focuses on TLS-1, but similar behaviour is seen for TLS-2 as well as other TLS. Furthermore, the simultaneous jumps of TLS-1 and TLS-2 in~\figref{fig:fig2}(a,b), while other defects remain unchanged, are consistent with coupling to a common electric environment.

\section{Timescales of memory effects}
\begin{figure*}[!t]
    \centering
    \includegraphics[width=\textwidth]{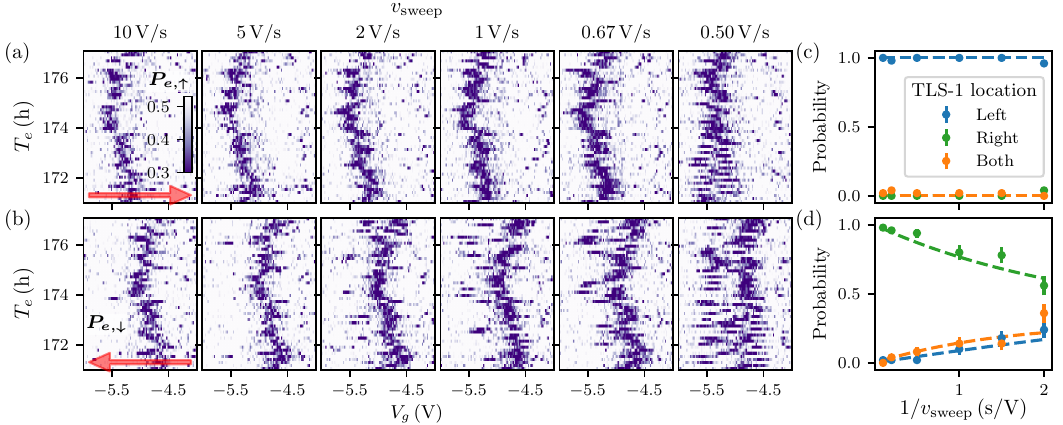}
    \caption{Results of interleaved experiments with different sweep rates $v_\text{sweep}$. The different sweep rates are shown in different columns, while (a) and (b) show $P_{e,\uparrow}$ and $P_{e,\downarrow}$, respectively. Sweep-rate dependence is not observed in (a), but is visible in (b), where TLS-1 appears at the lower-voltage location with higher probability for lower sweep rates. This is quantified in (c,d), where the probabilities of TLS-1 appearing in the left, right, or both locations are plotted. The dashed lines in (c,d) are fits to a model with CF symmetry point $V_\text{sym} \sim -4.6\,\text{V}$ and equilibration time $\tau_\text{eq} \simeq 5\,\text{s}$.}
    \label{fig:fig4}
\end{figure*}

Next we evaluate the timescales of the CF equilibration near its symmetry point by analysing the dependence of the probabilistic hysteresis observed in~\figref{fig:fig2}(f) on the voltage sweep rate. We perform experiments where we systematically vary the voltage sweep rate and interleave the experiments with the different rates. To keep the voltage resolution in the data constant, we add delays between individual repetitions of the pulse sequence such that the number of periods of voltage sweeps per experiment is constant. This interleaved experiment is then repeated for a period of $\sim 6\,\text{h}$. Data in the neighbourhood of TLS-1 are shown in~\figref{fig:fig4}(a,b).

For $P_{e,\uparrow}$, TLS-1 is predominantly located near $V_g\sim -5.4\,\text{V}$ and shows little dependence on sweep rate. In contrast, for $P_{e,\downarrow}$, TLS-1 is mainly located near $V_g\sim -4.7\,\text{V}$, but appears at the lower-voltage branch with increasing probability as the sweep rate is reduced. This behaviour suggests that the CF symmetry point lies above the lower-voltage branch, \(V_\text{sym}>-5.4\,\text{V}\). During increasing-voltage sweeps~[\figref{fig:fig4}(a)], the CF is therefore already in the equilibrium state associated with the left branch. During decreasing-voltage sweeps~[\figref{fig:fig4}(b)], however, the CF is initially polarised into the state associated with the right branch and can relax only after the sweep crosses $V_\text{sym}$. The probability of observing the left branch then depends on the time available for this relaxation. 

To evaluate the timescale of the CF equilibration, consider a model where after crossing the CF symmetry point at $V_\text{sym}$, the probability of the CF remaining in its initial state after time $t$ is $e^{-t/\tau_\text{eq}}$. This does not take into account the change in the equilibration time of the CF as the asymmetry energy is modified by the applied voltage and only serves as an approximation near the symmetry point of the CF. Let $P_{R,\downarrow}(V_g)$ denote the probability of the CF being in state $R$ for the decreasing voltage sweep at voltage $V_g$, where state $R$ corresponds to the state which leads to the TLS-1 resonance being on the right branch. We can describe this model as
\begin{equation}
P_{R,\downarrow}(V_g) =
\begin{cases}
    1, & V_g > V_\text{sym}, \\[4pt]
    \exp\!\left[-\dfrac{V_\text{sym}-V_g}{v_\text{sweep}\tau_\text{eq}}\right],
    & V_g \leq V_\text{sym}.
\end{cases}
\end{equation}
We vary $V_\text{sym}$ and $\tau_\text{eq}$ of the model to fit the experimental data and plot the best-fit results (dashed lines). We find $V_\text{sym} \sim -4.6\,\text{V}$ and $\tau_\text{eq} = 5.4\pm0.3\,\text{s}$, where the quoted uncertainty is the standard error obtained from the fit and does not account for model uncertainty. While approximate, this indicates remarkably long equilibration times, of the order of several seconds. Previously, such long timescales for charge relaxation have been only observed in other platforms such as single-electron transistors~\cite{pourkabirian_nonequilibrium_2014}. Depending on the physical processes governing the transitions these timescales may be affected by the temperature, which can be addressed in future work.

\section{Discussion}
In summary, we show that the TLS defect environment of a transmon qubit can retain memory of its electric-field history for several seconds. This memory appears as hysteresis in individual TLS resonance positions and is consistent with coupling to field-polarised charge fluctuators. The associated equilibration times are many orders of magnitude longer than the qubit lifetime and typical resonant TLS coherence times, showing that the microscopic defect bath contains long-lived internal degrees of freedom. 
Our method thus provides a route to unpicking microscopic fluctuator dynamics in superconducting circuits. Further extending our method, for example by using multiple gates~\cite{lisenfeld_mapping_2026}, scanning probe techniques~\cite{marius_in_situ_2025, banerjee2026coulombblockademicroscopicmaterial}, or automated feature extraction~\cite{agarwal_fast-tracking_2025}, can help further understand and disentangle interactions in the defect bath also in other devices suffering from TLS defects, such as semiconductor spin qubits~\cite{paladino_1f_2014,ye_2024_characterization_charge_fluctuators}. Furthermore, this understanding of interactions in the defect bath can be used to improve qubit stabilization and QEC protocols~\cite{chen2025scalablesitespecificfrequencytuning, lisenfeld_enhancing_2023, dane2025performancestabilizationhighcoherencesuperconducting,kim_error_2025}.

\section{Acknowledgements}
We acknowledge the support of the UK government Department for Business, Innovation, Science and Trade (BIST) through the UK National Quantum Technologies Programme, the Engineering and Physical Sciences Research Council (EPSRC) grant EP/Y0226371/1, and from the EURAMET European Metrology Partnership project 23FUN08 MetSuperQ where NPL is supported by UKRI grant number 10133632.
\bibliography{references}
\end{document}